\documentclass[journal,10pt,twocolumn,twoside]{IEEEtran}

\usepackage{cite}
\usepackage{graphicx}
\usepackage{longtable, ltcaption}
\usepackage{multicol}
\usepackage{comment}
\usepackage{algcompatible}
\usepackage{algorithm,algpseudocode}

\usepackage{mathtools}
\usepackage{subfigure}
\usepackage{color}
\usepackage{float}
\usepackage{multirow}
\usepackage{amsmath}
\usepackage{amsthm}
\usepackage[justification=centering]{caption}
\newtheorem{mydef}{Definition}

\theoremstyle{definition}
\newtheorem{exmp}{Example}

\begin{document}

\title{Ranking the Importance Level of Intermediaries to a Criminal using a Reliance Measure
}

\author{Pritheega~Magalingam*, ~Stephen~Davis,
        and~Asha~Rao,~\IEEEmembership{Member,~IEEE}
\thanks{P. Magalingam conducted this research while at the  School of Mathematical and Geospatial Sciences, RMIT University, Melbourne, GPO Box 2476, Melbourne, Victoria 3001, Australia. P. Magalingam is currently at the Advanced Informatics School, Universiti Teknologi Malaysia (UTM). e-mail: mpritheega.kl@utm.my}
\thanks{S. Davis and A. Rao are with School of Mathematical and Geospatial Sciences, RMIT University, GPO Box 2476, Melbourne, Victoria 3001, Australia. e-mail: stephen.davis@rmit.edu.au, asha@rmit.edu.au.}}

\maketitle

\begin{abstract}
Current research into discovering important intermediate nodes in a network suspected of containing criminal activity, is highly dependent on network centrality values. Betweenness centrality, for example, is widely used to rank the nodes that act as brokers in the shortest paths connecting all source with all end nodes in a network. However both the shortest path node betweenness as well as the linearly scaled betweenness can only show rankings for all the nodes in a network, and not for just a subset of source nodes. In this paper we explore the mathematical concept of pair-dependency of a source on intermediate nodes, adapting the concept to criminal relationships and introducing a new source-intermediate reliance measure. To illustrate our measure, we apply it to rank the nodes in the Enron email dataset and the Noordin Top Terrorist networks. We compare the reliance ranking with Google PageRank, Markov centrality as well as betweenness centrality and show that a criminal investigation using the reliance measure, will lead to a different prioritisation in terms of possible people to investigate. While the ranking for the Noordin Top terrorist network  yields more extreme differences than the Enron email transaction network, in the latter the reliance values for the set of finance managers immediately identified another employee convicted of money laundering.
\end{abstract}

\begin{IEEEkeywords}
Shortest path, betweenness, intermediate node, reliance
\end{IEEEkeywords}

\begin{center} \bfseries EDICS Category: FOR-OTHS, OTH-BGDT \end{center}

\section{Introduction}
\label{Introduction}

A number of methods have been proposed for ranking the important nodes in a network. One such measure, the betweenness centrality attributes importance to intermediate nodes in a path as these nodes are necessary to retain the flow within the network \cite{brandes2008variants}. However, given a node of interest, $u$,  betweenness centrality cannot rank the nodes important only to $u$, instead ranking the nodes with regards to ensuring flow in the network as a whole. In this paper, we present a new measure, the reliance measure, that achieves this aim of ranking nodes based on their importance to a given node. Knowing one suspect, our reliance measure would allow the creation of an ordered list of trusted connections. Thus this method of ranking nodes would enable an investigator to prioritize her search for criminal connections. Our research has the specific aim of aiding the investigation of money laundering crimes, where it is often difficult to identify influential entities with respect to the main source of illegal money by  using only data mining techniques \cite{zhang2003, le2010, tang2005, gao2007}. 

To show the versatility of our measure, we apply it to rank  nodes in the Enron dataset \cite{ISI:2009} and the Noordin Top Terrorist networks \cite{everton2012}. Our reliance measure is able to show clear differences in the ranking of particular Enron employees as well as within members of the terrorist network making it easy to pick important people for further investigation. In contrast, the betweenness centrality  ranking of the same nodes does not indicate any nodes of interest.  Indeed, in our experiment, using the set of Enron finance managers, the reliance ranking highlights an employee who was not a finance manager, but was convicted of money laundering.

The most well known method of ranking intermediate nodes in a network is betweenness centrality. The betweenness centrality of a node is a measure that computes the number of geodesics (shortest paths) going through that node. In a network, the node that appears the most number of times in the shortest paths linking every pair of nodes or components acts as a broker or intermediary \cite{scott2012, borgatti2006}, and has the highest betweenness centrality value. Researchers,  including Xu and Chen \cite{xu2005criminal}, and Morselli \cite{morselli2010assessing}, have used betweenness centrality to identify a gatekeeper in criminal networks while others, \cite{Ferrara2015,drezewski2015application}, have used it  to identify potential persons of influence in these networks.  Cantanese et al. \cite{catanese2013}, for example, utilise multiple network metrics in their log analysis tool to identify key members in a criminal network, in particular, using betweenness centrality to show the communication control of one node over other nodes. In all of this research betweenness centrality is directly used to identify an important node that provides the communication path to many other nodes. However, using betweenness centrality and/or degree centrality is insufficient when identifying important nodes in a money laundering network.

Money laundering networks are inherently different to other criminal networks, in that they may have many members who are important to a suspect but the nature of the crime dictates that these members are hidden or inactive (not possessing a high centrality value).  Research, for example \cite{farasat2015} and \cite{lindelauf2013cooperative}, shows that the node with highest betweenness centrality could also have the highest closeness as well as highest degree centrality. Both of these latter attributes point to nodes that are highly visible. Thus, using betweenness centrality alone is an impractical way of picking an important person in a money laundering network. In addition, in money laundering, a suspect could rely more on a person at quite some distance from them. We address both of these issues in this paper.

Our research, thus, starts with the idea of dependence. Researchers have approached the task of finding influential nodes  by combining  the concept of dependence with betweenness, where the dependence of  node \textit{i} on node \textit{j} in a network reflects in some way the influence of node \textit{j} on node \textit{i} \cite{kourtellis2013}. Dependency can be said to exist when there is information flow between a source and a destination. For example, while identifying important people in the Enron email dataset, Shetty et al. \cite{shetty2005} characterised email link dependency in various ways; an email was marked as dependent on another when both the emails appeared within the same time frame, a major part was copied from one email to another, and the received email was then forwarded, or when the links in the email were based on a particular event. The authors based these characterisations on intuition and domain knowledge which makes it difficult to interpret the  relationships, discovered by them, between individuals. 

In this paper, we propose a new dependency formula to calculate the dependence of a source node on an intermediate node. We name this dependency, \textit{reliance}. The reliance formula is used as a tool to build a source-intermediate node reliance measure algorithm that can measure the suspicious sub-network that contains the criminals and/or suspects and their associates. The reliance values are used to rank the nodes upon which a criminal relies. This could then allow the identification of nodes that are relied on by a group of known suspects and thus progress the criminal investigation.  

The commonly deployed centrality measures, including betweenness centrality, compute the central value of a node by summing the values over all source nodes in a network. Our reliance formula focuses on a particular source node and on the intermediate nodes in the shortest paths from this particular source node to every possible end node in the network. Our interest is in the relationship between a specific source node (either a criminal or suspect) and its intermediate nodes because an intermediary carrying information tends to hide illicit activity either on purpose or inadvertently \cite{campbell2013}. Campbell \cite{campbell2013} mentions in his paper that organized crimes such as money laundering, smuggling, and trafficking of antiquities can be linked to criminals who have multiple associations. Thus, a mafia member who acts as an intermediary in a mafia endeavor could escape criminal charges, as, most commonly, identification is done by computing the central value of a node; summing over all possible source nodes in the network, and the intermediary may not stand out in this manner.

We illustrate the use of our reliance measure by applying it to the Enron email transactions \cite{ISI:2009}, obtaining a reliance sub-network  to aid a money laundering criminal investigation. History shows that some senior officials at Enron were involved in accounting irregularities \cite{bbc2002RiseandFall} leading to an investigation of various internal business and financial activities including the partnerships set up by some finance managers. This was followed in early January 2002 by an inquiry into other criminal activities involving many Enron executives \cite{bbc2002}, and finally to the conviction of 10 individuals of money laundering \cite{Kathleen2003}. In \cite{MagalingamP2014}, we built a shortest path network search algorithm (SPNSA) using shortest paths combined with two centrality measures; eigenvector centrality and  betweenness centrality. We applied this SPNSA \cite{MagalingamP2014, Magalingam20151} to the Enron dataset, extracting a sparse and more manageable network of people for further criminal investigation. The method in \cite {Magalingam20151} began with the identification of a list of suspects (the algorithm feed) to extract a sub-network. Here we use the new reliance measure proposed in this paper to calculate the reliance values and rank each intermediate node of each suspect in this extracted network.  For the purpose of illustration, we compare our reliance ranking with three other ranking measures; betweenness, Google PageRank and Markov centrality. 

The rest of the paper is organised as follows: section \ref{preliminary} contains the definitions and mathematical terms used, while section \ref{ourapproach} describes the proposed method of calculating the reliance of source nodes on the intermediate nodes. In the section following this, we compare the pair-dependency formula given by Freeman \cite{freeman1977}, Brandes \cite{brandes2001},  and Geisberger et al. \cite{geisberger2008} with our reliance method as well as quantifying the difference between the reliance measure and Geisberger et al.'s formula. Section \ref{datasets} describes the two datasets; the Enron email transactions and the Noordin Top Terrorist networks used in our experiments. In section \ref{sec:casestudy} we apply the reliance measure to these two datasets to show the difference in ranking between the reliance measure and the Brandes and Geiseberger et al. measures. as well as comparing the reliance rankings with those found using both the Markov walk-path and the Google Pagerank algorithms. Section \ref{IdentifyingCrimePriorityNodesMain} details the prioritisation of nodes for the purpose of a criminal investigation. Finally we give the conclusion.

\section{Preliminaries}
\label{preliminary} 

Since our reliance formula is closely related to the betweenness centrality measure, we will start with the necessary graph-theoretic terminology \cite{harary1969}, and the three main definitions of betweenness; Freeman \cite{freeman1977}, Brandes \cite{brandes2001} and Geisberger et al. \cite{geisberger2008}. For more details see Brandes \cite{brandes2008variants} and Newman \cite{newman2010}.

Let $G = (V,E)$ be a graph where $V=\left\{v_1,v_2,...\right\}$ is the set of vertices  (also called nodes) and $E=\left\{e_1,e_2,...\right\}$ the set of edges  (representing the connections between the vertices), with the total number of vertices and edges given by $\left\vert{V}\right\vert$ = $n$ and $\left\vert{E}\right\vert$ = $m$, respectively. An edge that has the same start node and end node is called a self-loop or a loop. If more than one edge is associated with a pair of nodes, these are called multiple edges. For our purpose, we exclude all self-loops and consider multiple edges as one edge.

A path is a sequence of edges that connects multiple nodes \cite{newman2010}. Given a path $(s,t)$, we call $s$ the source node and $t$ the destination, end node or target node. In between the source and the target, lies the alternating sequence of nodes and edges, for instance, $s,e_1(s,v_1),v_1,e_2(v_1,v_2),v_2,...,e_t(v_{i},t),t$, that make up the path $(s,t)$. Here $e(u,v)$ denotes the edge connecting nodes $u$ and $v$. In the graph $G$, the length of an $(s,t)$-path is the number of edges it contains, and the distance, $\mu(s,t)$, from $s$ to $t$ is defined as the minimum length of any $(s,t)$-path if one exists and undefined otherwise \cite{brandes2008variants}. Let the number of shortest paths from $s$ to $t$ be given by $\sigma_{st}$, and let $\sigma_{st}(v)$ be the number of shortest paths from $s$ to $t$ that pass through $v$. 

\begin{mydef}[Freeman \cite{freeman1977}]
The pair-dependency $\delta_{st}(v)$ of a pair of nodes $s$ and $t$ on an intermediate node $v$ is the proportion of the shortest paths from $s$ to $t$ that contain $v$, that is:

\begin{equation}\label{pair-dependence}
\delta_{st}(v) = \frac{\sigma_{st}(v)}{\sigma_{st}}
\end{equation}
\end{mydef}

The betweenness centrality of $v$  is then the sum of all such pair-dependencies \cite{freeman1977}:

\begin{equation} \label{BC_Freeman}
BC(v) = \sum_{{s \neq v \neq t \in V}}\delta_{st}(v)
\end{equation}

In 2001, Brandes \cite{brandes2001} introduced an algorithm for computing betweenness centrality in a network.

\begin{mydef}[Brandes' Algorithm \cite{brandes2001}] 
The dependence of $s$ on an intermediate node $v$ is given by:

\begin{equation}\label{Brandes}
\delta_{s*}(v) = \sum_{{w:v \in P_{s}(w)}}\frac{\sigma_{sv}}{\sigma_{sw}} (1+ \delta_{s*}(w)) 
\end{equation}

Here $\{{w:v \in P_{s}(w)}\}$ is the set of all nodes $w$ where $v$ is an immediate predecessor of $w$ in a shortest path from $s$ to $w$, that is $v \in P_{s}(w)$.
\end{mydef}

According to Brandes \cite{brandes2001}, $\delta_{s*}(v)$, the dependence of $s$ on $v$ is positive, that is $\delta_{s*}(v)$ $>$ 0 only when $v$ lies on at least one shortest path from $s$ to $t$ and on any such path there is exactly one edge $\{v,w\}$ with 
$v \in P_{s}(w)$. The Brandes' algorithm has been used among other things, to identify the highest betweenness of an edge that separates two communities \cite{tyler2005}.


In 2008, Geisberger et al. applied a linear scaling to Brandes' algorithm by introducing the length function (unit edge weight) \cite{geisberger2008} thus obtaining a better approximation for betweenness centrality on large networks without overestimating the values for the nodes near a pivot or parent node which was the case with Brandes' algorithm \cite{brandes2001}. 

\begin{mydef}[Geisberger et al. \cite{geisberger2008}]
Given a shortest path from source $s$ to a node $w$ with node $v$ a predecessor of $w$ on this path, the length function is the ratio of the distance $\mu(s,v)$ of $v$ from  $s$ to the distance of $w$ from $s$, $\mu(s,w)$. Thus, the dependence of $s$ on an intermediate node $v$ changes to:

\begin{equation}\label{Geis}
\delta_{s*}(v) = \sum_{{w:v \in P_{s}(w)}}\frac{\mu(s,v)}{\mu(s,w)}  [\frac{\sigma_{sv}}{\sigma_{sw}} \times (1+ \delta_{s*}(w))]
\end{equation}
\end{mydef}

While Brandes' algorithm gives good exact results for small networks, often it is not possible to get exact results in reasonable running time for large networks. With this change,  Geiseberger et al., could apply betweenness centrality to real world situations such as choosing improved highway-node routings \cite{geis2008, geisberger2008}.

\section{The reliance measure}
\label{ourapproach}
As mentioned before, the betweenness centrality algorithms, given in the previous section, cannot calculate dependence for particular source nodes. In our proposed reliance formula, we start with a specific set of source nodes $\left\{s_{1},s_{2},...\right\}$ and calculate the reliance (a.k.a. ``dependency") value for each intermediate node $v$ that occurs on paths starting from each $s_{i}$ to all possible end nodes $t$. Thus, we consider all shortest paths from a particular source $s$ to all possible end nodes $t$ where $s \neq t$, and the source node $s$ comes from a specific set of interest, rather than all possible nodes as is done in equations (\ref{BC_Freeman}), (\ref{Brandes}) and (\ref{Geis}). 

 In the first part of our formula, we calculate the proportion of the shortest paths linking source $s$ to all nodes $t$, that contain $v$ (equation (\ref{pair-dependence})), and call it the \emph{importance rate} $IR_{(s,t)} (v)$. We name the pair dependency of equation (\ref{pair-dependence}) as the importance rate of $v$ as it indicates how often the node $v$ is relied on by $s$ to complete a path from  $s$ to a destination $t$ in proportion to all paths from $s$ to $t$. 

The second part of our formula gives a trust value that the source $s$ places on $v$ to pass messages to any $t$ along the shortest paths. This trust value allows the measuring of trust between nodes that are far away from each other. The idea that someone has to trust a person much further down the chain of communication is an important concept in money laundering. 

Verbiest et al. \cite{verbiest2012} in their research incorporating path length and trust aggregation declare that the shorter the distance from source $s$ to $v$, the more the source $s$ trusts $v$. Similarly, De Meo et al. \cite{de2012} design an algorithm to compute edge centrality using k-path length with the assumption that the influence between two nodes reduces when the distance between them increases. Indeed, a common way to start a criminal investigation process is to identify the closest node to a criminal or source because the shorter the distance from the source, the higher the chances the node is the source's subordinate \cite{malm2008}.

This is not the case with money laundering. We claim instead that the further away an intermediate node is from a source node,  the more the source needs to trust that intermediate node to pass on a message, in the case of money-laundering. One of the stages of money laundering is the layering of money \cite{AUSTRAC:2008}  where the money is moved to different channels or banks through intermediaries such as corporations or trusts. The criminals are consequently connected to multiple bank accounts that are, in turn, connected to other people. The intermediaries are different people in the middle who help to channel the money to these different destinations so as to obscure its illegal source.

Thus, the layering process in money laundering involves multiple transactions of money through various channels \cite{AUSTRAC:2008, AUSTRAC:2011}.  In such a money layering process, money below the threshold is distributed to different financial institutions or accounts. This contributes to the growth in the length of the paths that are used to transport the money from a source to a destination. The source uses longer and longer sequences of channels to divide and distribute smaller amounts of money making it more and more difficult for law enforcement authorities to identify the influential people \cite{AUSTRAC:2008, AUSTRAC:2011}. Clearly, this upholds our claim that in a crime such as money laundering, the source of the illegal funds has to place more trust on a person further down the chain. 

The trust formula defined below is able to measure this increased trust.

\begin{mydef}[Trust]
Given a graph $G = (V,E)$,  $s$ a source node and $v$ an intermediate node on some path $(s,t)$ from $s$ to an end node $t$, the trust of $s$ on $v$, relative to the path $(s,t)$ denoted by $T_{(s,t)}(v)$, is given by:

\begin{equation}
T_{(s,t)}(v) = \frac{\mu(s,v)}{\mu(s,t)}, \mbox{ for $t \neq s, s \neq v\neq t$}
\end{equation}
where $\mu(s,u)$ is the minimum distance (number of edges) from $s$ to $u$ along the path $(s,t)$, $u \in V$, if one exists and undefined otherwise.
\end{mydef}

We illustrate the trust concept using a representative money laundering network.

\begin{exmp}
In Figure \ref{fig:ASampleGraph}, let 1 be the source node and 8 the destination or end node. A path from source 1 to destination 8 is:

\begin{center}
1 $\rightarrow$ 2 $\rightarrow$ 3 $\rightarrow$ 5 $\rightarrow$ 7 $\rightarrow$ 8
\end{center}

\begin{figure}[ht]
\DeclareGraphicsExtensions{.pdf,.png,.jpg,.tiff}
\centering
\includegraphics[width=3.1 in]{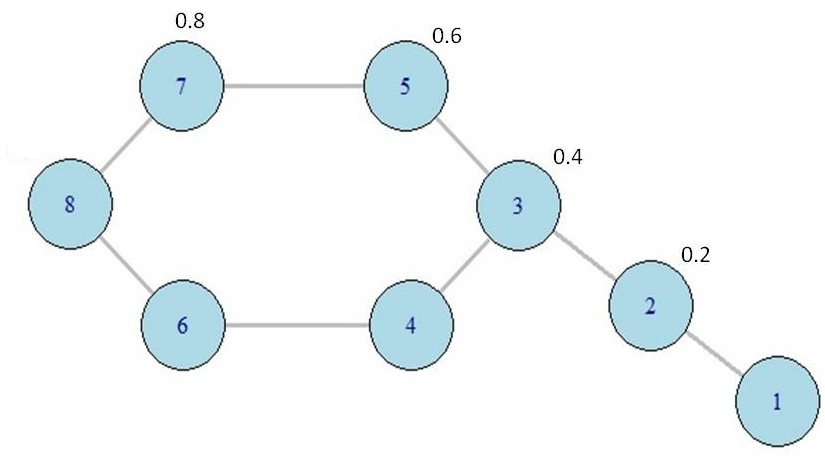}
\captionsetup{singlelinecheck=off,justification=justified}
\caption{\small \textbf{A representation of a small money laundering network.} Node 1 could be considered the source and node 8, the destination of the money.  The floating-point numbers near each node represent the trust placed by source node 1 on each intermediate node from the set $\left\{2, 3, 5, 7\right\}$ in the path 1 $\rightarrow$ 2 $\rightarrow$ 3 $\rightarrow$ 5 $\rightarrow$ 7 $\rightarrow$ 8.}
\label{fig:ASampleGraph}
\end{figure}

Suppose Figure \ref{fig:ASampleGraph} is a representation of a money laundering network and node 1 was a money laundering suspect. An example of when this might be the case is the layering of illegal money. Hence consider Figure 1 as representing the layering of illegal money within a money laundering syndicate, with node 1 the source of the money being laundered. 

In such a scenario, the source node (node 1 in Figure \ref{fig:ASampleGraph}) needs to trust node 7 (at distance 4) more than node 3 (at distance 2) to pass on the message (or money) to node 8. This difference is reflected in the trust value, with $T_{(1,8)}(7)$ = $\frac{4}{5}$ while $T_{(1,8)}(3)$ = $\frac{2}{5}$.

\end{exmp}

 However, it is insufficient to just have a trust value, as the number of paths that exist from a source and a destination would also play a role, as would the number of paths that the particular intermediate node lies on, to work out the reliance of the source on this intermediary.  Thus, we incorporate the importance rate and the trust value of an intermediate node, to get the reliance of source $s$ on an intermediate node $v$.   

We focus on a particular source, $s$, that we call a suspect and our motive is to calculate the reliance of this suspect on the intermediate nodes that reside along all paths $(s,t)$.

\begin{mydef}[Reliance]
Given a graph $G = (V, E)$, with $|V|=n$, a source node $s$ and an intermediate node $v$ on some path from $s$ to a particular end node $t$, given by $(s,t)$, the reliance of $s$ on $v$ along the path $(s,t)$, $r_{(s,t)}(v)$ \mbox{ for $v \in (s,t), t \neq s, s\neq v \neq t$}, is the product of the importance rate $IR_{(s,t)}(v)$ and the trust $T_{(s,t)}(v)$:

\begin{equation}\label{reliance_1}
r_{(s,t)}(v) =\delta_{st}(v) \times \frac{\mu(s,v)}{\mu(s,t)}
\end{equation}

Equation (\ref{reliance_1}) gives the reliance of $s$ on $v$ along a particular path $(s,t)$. Hence, the total reliance of source $s$ on $v$ over all paths from $s$ to all possible end nodes $t$, is:

\begin{equation}\label{reliance_all}
R_{s}(v) = \sum_{\{t: v \in (s,t), t \neq s, s \neq v \neq t\}} \frac{r_{(s,t)} (v) }{(n-2)}
\end{equation}
\end{mydef}

Since $|V|=n$, that is, there are $n$ vertices in the graph, and since the start node $s$ and the intermediate node $v$ are fixed and $r_{(s,t)}(v)$ is taken for all $t$, where $t \neq s, s\neq v \neq t$, there are $(n-2)$ possible values for $t$, thus, we normalise the reliance value, $R_{s}(v)$ with $(n-2)$.

\section{Comparison between different dependency techniques}
\label{sec:test}

We compare our reliance measure to the betweenness calculations given by Freeman \cite{freeman1977} (equation  (\ref{BC_Freeman})), Brandes \cite{brandes2001} (equation (\ref{Brandes})) and Geisberger et al. \cite{geisberger2008} (equation (\ref{Geis})), and show that even though all four formulae use the dependency given by equation (\ref{pair-dependence})  with, in addition,  Geisberger et. al's formula using a version of the length function, there are fundamental differences in these formulae and that the reliance formula is best able to do the job of ranking nodes with respect to a particular source.

\begin{exmp}\label{1stComp} \emph{Comparison of reliance with Freeman \cite{freeman1977}, Brandes \cite{brandes2001} and Geisberger et al. \cite{geisberger2008} betweeneness.} We use the graph in Figure \ref{fig:ASampleGraph} to compare the  reliance of node 1 on  intermediate nodes of the network with the values assigned to the intermediate nodes by the three dependency measures, as given by equations (\ref{BC_Freeman}), (\ref{Brandes}) and (\ref{Geis}) respectively. The results are shown in Figure \ref{fig:Graph1}.

\begin{figure}[ht]
\DeclareGraphicsExtensions{.pdf,.png,.jpg,.tiff}
\centering
\includegraphics[width=3.5 in, height=2.2 in]{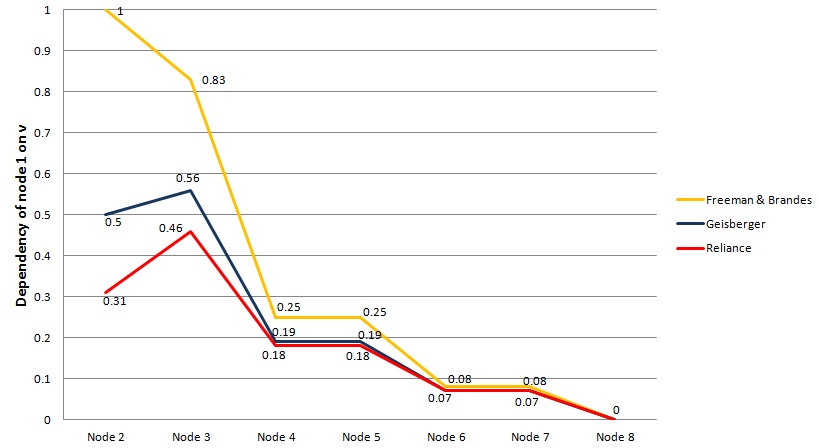}
\captionsetup{singlelinecheck=off,justification=justified}
\caption{\small \textbf{The result of applying the different dependency formula and the reliance formula} on the network shown in Figure \ref{fig:ASampleGraph}. Here, for the purpose of comparison we choose node 1 as the source as  and take all shortest paths from this source node to all possible end nodes. The value of reliance of node 1 on the intermediate nodes is compared to the dependency values given by  eqns. (\ref{BC_Freeman}), (\ref{Brandes}) and (\ref{Geis}).}
\label{fig:Graph1}
\end{figure}

The reliance (dependency) of node 1 on itself is zero. Similarly, the reliance and dependence value of node 1 on node 8 are zero because there is no shortest path from source 1 to any possible end node that contains 8 as an intermediate node. The biggest difference that we can see from the graph in Figure \ref{fig:Graph1} is in the dependence of node 1 on node 3. Both Freeman and Brandes  give the dependence value of node 3 as less than that of node 2, whereas Geisberger et. al  puts the dependence value of node 3 as more than that of node 2. Similarly the reliance measure has node 1 relying more on node 3  than on node 2. The network in Figure \ref{fig:ASampleGraph} is undirected and node 3 is at the crucial position of separation between two sets of nodes. Removing node 3 cuts the information flow from node 1 and node 2 to the other nodes. Thus, based on the position of the nodes, the dependence and the reliance value of node 1 on node 3 should be intuitively higher. Both Geisberger et. al's technique and reliance measure reflect this.  
\end{exmp}

Even though the ranking of the nodes in the graph in Figure \ref{fig:ASampleGraph} is the same for Geisberger and reliance, the technique by Geisberger et. al is different from our reliance formula as the reliance calculation focuses on the reliance of a specific source on other nodes whereas, in general, Geisberger et. al's formula gives betweenness by considering all sources. Consequently, as the graph gets bigger, the  estimation by Geisberger et. al 's technique gets much larger than our reliance measure; something that is clearly illustrated in the next example. 

\begin{exmp} \emph{Difference between the reliance formula and Geisberger et. al's formula}: We use a slightly larger graph (Figure \ref{fig:SampleGraph2}) to illustrate the growth in dependence using Geisberger's formula when compared to the reliance calculation.

\begin{figure}[ht]
\DeclareGraphicsExtensions{.pdf,.png,.jpg,.tiff}
\centering
\includegraphics[width=1.8in]{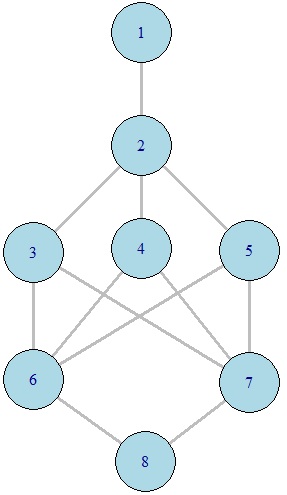}
\caption{\small \textbf{A sample graph.} This sample graph is used to compute the dependency of node 1 on other nodes.}
\label{fig:SampleGraph2}
\end{figure}

Figure \ref{fig:SampleGraph2} shows a network with 8 nodes. We calculate the dependency of node 1 on other nodes using Geisberger et al.'s technique (equation (\ref{Geis})).

\begin{small}
\begin{flushleft}
\begin{eqnarray}
\delta_{s=1*}(v) =& \hspace*{-1.5cm} \sum_{{w:v \in P_{s}(w)}}\frac{\mu(s,v)}{\mu(s,w)}  [\frac{\sigma_{sv}}{\sigma_{sw}} \times (1+ \delta_{s*}(w))] \nonumber\\
=& \hspace*{-1cm}  \left\{\sum_{\{t:v \in (s,t), t \neq s, s\neq v \neq t\}}\left[\frac{\mu(s,v)}{\mu(s,t)} \times \delta_{st}(v) \right] \right \} + \nonumber\\
& \hspace*{-1cm} \left\{2\left\{\frac{\mu(s,2)}{\mu(s,6)}[\frac{\sigma_{s2}}{\sigma_{s6}}] \right\} +  2\left\{\frac{\mu(s,2)}{\mu(s,7)}[\frac{\sigma_{s2}}{\sigma_{s7}}] \right\} + 5\left\{\frac{\mu(s,2)}{\mu(s,8)}[\frac{\sigma_{s2}}{\sigma_{s8}}] \right\}\right \} \nonumber\\
\hspace{-4em} =& \hspace{-1em} R_{s}(v) + 2\left\{\frac{\mu(s,2)}{\mu(s,6)}[\frac{\sigma_{s2}}{\sigma_{s6}}] \right\} + 2\left\{\frac{\mu(s,2)}{\mu(s,7)}[\frac{\sigma_{s2}}{\sigma_{s7}}] \right\} \nonumber\\
& \hspace*{-1cm} + 5\left\{\frac{\mu(s,2)}{\mu(s,8)}[\frac{\sigma_{s2}}{\sigma_{s8}}] \right\}
\end{eqnarray}
\end{flushleft}
\end{small}
\end{exmp}

\vspace{5mm}


Thus equation (\ref{Geis}) over-estimates the dependency by $\left\{2\left\{\frac{\mu(s,2)}{\mu(s,6)}[\frac{\sigma_{s2}}{\sigma_{s6}}] \right\} +  2\left\{\frac{\mu(s,2)}{\mu(s,7)}[\frac{\sigma_{s2}}{\sigma_{s7}}] \right\} + 5\left\{\frac{\mu(s,2)}{\mu(s,8)}[\frac{\sigma_{s2}}{\sigma_{s8}}] \right\}\right\}$  due to the repetitive dependency calculation of the source node on nodes 6, 7 and 8. Thus the direct use of the Geiberger et al. formula for the purpose of calculating reliance would result in an overestimation of the dependence values of certain nodes, a fact that becomes clear in section \ref{sec:casestudy}. While Geiseberger et al.'s formula is a good means of estimating values for large networks, our purpose here is to further refine a possible network for criminal investigation, and rank nodes for a specific set of source nodes.

\section{Datasets}
\label{datasets}

We use two different datasets in our experiments - the Enron dataset and the Noordin Top Terrorist Dataset. 

\subsection{The Enron Network}
\label{EnronDataset}

In \cite{Magalingam20151}, we first divided the Enron email dataset, which consist of 1,887,305 email transactions into two groups; emails that were sent using only the fields `TO/CC'  and those emails that also contained `BCC' recipients. Our SPNSA implementation in \cite{MagalingamP2014} was based on the number of BCC recipients needed to identify a trust network, where the focus was on emails that contained one or two recipients bcc-ed. By their very nature, `BCC 'email transactions contain recipients that are kept secret \cite{fox2012}. In \cite{Magalingam20151}, our experimental results showed that the undirected BCC email transactions have the most number of criminals in the shortest paths network, thus further analysis was conducted using only the BCC email transactions.  In this paper, both the `TO/CC' and `BCC' undirected email transaction networks are used to produce the reliance sub-networks for all suspects. Through dividing the email transactions, we are able to compare the important nodes that a suspect relies on, whether or not the connection (email) is kept secret (bcc-ed). To form an undirected network, we make the broad assumption that an email sent from A to B or from B to A implies an \textit{undirected} relationship between A and B.

The majority, 87.3\%, of email transactions in the Enron dataset, use the fields `TO/CC', while 12.7\% are `BCC' email transactions (i.e. they contain bcc'ed recipients). The email transactions in this dataset comprise of external and internal email addresses where the external email addresses are those that do not have `@enron' whereas the internal emails do. The email addresses are the nodes and we give importance to the sequence of emails exchanged by the nodes rather than the content of the emails.  As in \cite{MagalingamP2014, Magalingam20151}, here too the sets of source nodes, consisting of Enron finance managers as well as a few others who held top posts, were selected based on the possibility of having been involved in money laundering and are henceforth called suspects, and used as feed to the SPNSA algorithm \cite{MagalingamP2014}. We run our experiments using two sets of suspects; the first set of suspects consisting of the Enron finance managers \cite{anderson2003} while the second, larger, set comprises of all managers \cite{anderson2003} including the finance managers. Both sets were collected from a report on the chronology of events related to the collapse of Enron \cite{anderson2003}.

A manager may be indexed by more than one node if he or she has more than one email account. The two email transaction groups, `TO/CC' and `BCC', have distinct ID sets for the different email addresses, designed as such because, somewhat surprisingly, some nodes that exist in the `BCC' group do not exist in the `TO/CC' group.  The network formed using the `BCC' email transaction subset has 19,716 nodes and 65,532 edges while the network formed using the `TO/CC' email transaction subset has 26,027 nodes and 252,863 edges. All self-loops and multiple edges have been removed from these networks.

\subsection{The Noordin Top Terrorist Network}
\label{TerrorDataset}

The second dataset that we use to compare rankings is the Noordin Top terrorist network \cite{everton2012}. This dataset is small and consists of different types of connections. The first group of connections gives the terrorists' affiliations such as terrorist organisations, educational institutions, business and religious institutions. The second group contains relationship information such as classmates, kin, friends with the third group comprising of individuals that provided logistical support or participated in training events, terrorist operations, and meetings \cite{everton2012}. We take 2 different subsets from this dataset for our analysis, the  terrorist-friendship and terrorist-classmates.

In our experiment, we rank the terrorists using source-intermediate reliance value. Since it is a small network, we do not extract any terrorist sub-network using SPNSA \cite{MagalingamP2014} instead using the terrorist-friendship and terrorist-classmates networks as they are.

\section{Ranking important nodes using the Reliance Measure}
\label{sec:casestudy}

As we saw in section \ref{sec:test}, even in small networks, there is a difference between the dependence rankings given by the betweenness centrality methods, and the reliance measure. In this section, we illustrate further the need for our reliance measure. 

In subsection \ref{CompareBCandRelianceonEnronNetwork}, for the purpose of illustration, we compare the ranking of the nodes in the Enron `BCC' network using betweenness centrality and the reliance formula.  We also compare the reliance value ranking with the rankings obtained using a Markov centrality score and the Google PageRank method. 

Random walks are used to calculate Markov centrality scores \cite{white2003}, with the centrality score of a node being calculated by first taking the average path length of a random walk starting at that node and arriving (for the first time) at some other node, and then averaging over all other nodes. Markov centrality of the node is then the inverse of the average distance between it and every other node \cite{white2003, SANTAPackage}. The PageRank algorithm, used by Google to rank important web pages, uses the assumption that a page is important when it is linked by many pages or if it is linked to many other important pages\cite{kamvar2004, brin1998}. The mathematical equivalent of this concept is the eigenvector centrality measure \cite{kamvar2004}.  

For the Noordin Top terrorist network, in section \ref{CompareBCandRelianceonNoordinTopTerroristNetwork} we use these measures to rank a terrorist, his friends and classmates to see if there are any differences between reliance ranking to other measures.

\subsection{Comparing reliance node ranking with betweenness centrality, Markov centrality and PageRank on the Enron network}
\label{CompareBCandRelianceonEnronNetwork}

We use only the Enron BCC Network for this part of the experiments. We compare the node rankings resulting from using the betweenness centrality proposed by Brandes and Geisberger et.al with the results produced using the reliance measure on particular subsets of the Enron `BCC' network.  Since betweenness centrality measures the dependency of all source nodes on $v$, we decided that a comparison was best done by comparing the betweenness value of a node $v$ with its combined reliance value $\sum_{s \in S} R_s(v)$, obtained by taking the sum of the reliance of all $s$ in the subset $S$, where $S$ is the set of all finance managers (respectively, all managers), on this $v$ in the Enron finance manager (respectively, manager) BCC sub-networks. The nodes $v$ are ranked based on the decending order of betweenness centrality (Brandes' and Geiseberger et. al) values. 

We start our comparisons with the finance manager sub-network extracted from the BCC network. Five employees worked as finance managers in Enron  between the years 1990-2001 and were used as the SPNSA's \cite{MagalingamP2014}  feed.  These finance managers \cite{anderson2003} were Andrew Fastow (686, 687), Sherron Watkins (16929), Ben Glisan (1369), Rick Causey (15077) and Jeff McMahon (8071) and the sub-network extracted, the Enron Finance Manager BCC sub-network, had 30 nodes and 53 edges. As mentioned before, the betweenness centrality measures by Brandes and Geisberger et al. calculate the centrality value of $v$ using all source nodes. To compare the reliance value with the betweenness centrality measures, we take the total reliance of all finance managers, $F = \{ 686, 687, 16929, 1369, 15077, 8071 \}$, on each $v$ in the Enron finance manager  BCC sub-network. Thus, the total reliance value for each possible intermediate node $v$ is $\sum_{s \in F} R_{s}(v)$. This compares well with the betweenness centrality value for $v$ in the case of both Brandes and Geiseberger et al.  which is $\delta_s* (v)$  for all $s$ in the sub-network, and not just $s \in F$.

In addition, we normalise the total reliance as well as the betweenness centrality values obtained using Brandes and Geisberger et al. by dividing each node's value with the maximum value in each set respectively.

\begin{figure}[ht]
\DeclareGraphicsExtensions{.pdf,.png,.jpg,.tiff}
\centering
\includegraphics[height=1.6 in,width=3.5 in]{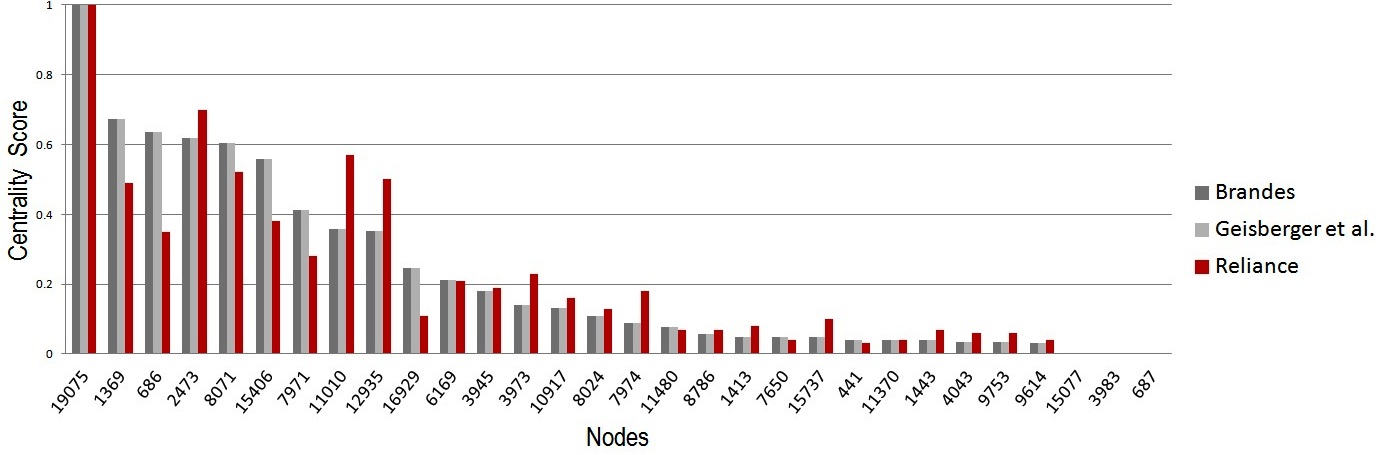}
\captionsetup{singlelinecheck=off,justification=justified}
\caption{\small \textbf{The Enron finance manager BCC sub-network node ranking.} The total reliance of all finance managers on each $v$ in the Enron Finance Manager BCC sub-network is used for this comparison. The intermediate nodes $v$ are ranked based on the descending order of betweenness centrality (Brandes) values. There is a clear difference in node ranking between the total reliance measure and betweenness centrality measures. For example, node 2473 is more heavily relied on by the finance managers than 686, where as both the betweenness centrality measures rank these two the other way around.}
\label{fig:BCValueFM}
\end{figure}

The comparison of betweenness centrality with reliance node ranking for the nodes relied on by the Enron finance managers (in the Enron Finance Manager BCC network) is depicted in Figure \ref{fig:BCValueFM}. 


In Figure \ref{fig:BCValueFM}, all three measures identified the same node as having the highest value. The interesting point is to note that some of the nodes (for example, 11010, 12935 etc.) ranked lower by the betweenness centrality measures, are ranked as being more important by the reliance measure. Node 11010 (lfastow@pop.pdq.net), ranked third by reliance, belongs to Lea Fastow, who was convicted of money laundering \cite{MagalingamP2014, Kathleen2003}. 

\begin{figure}[ht]
\DeclareGraphicsExtensions{.pdf,.png,.jpg,.tiff}
\centering
\includegraphics[height=1.8 in,width=3.5 in]{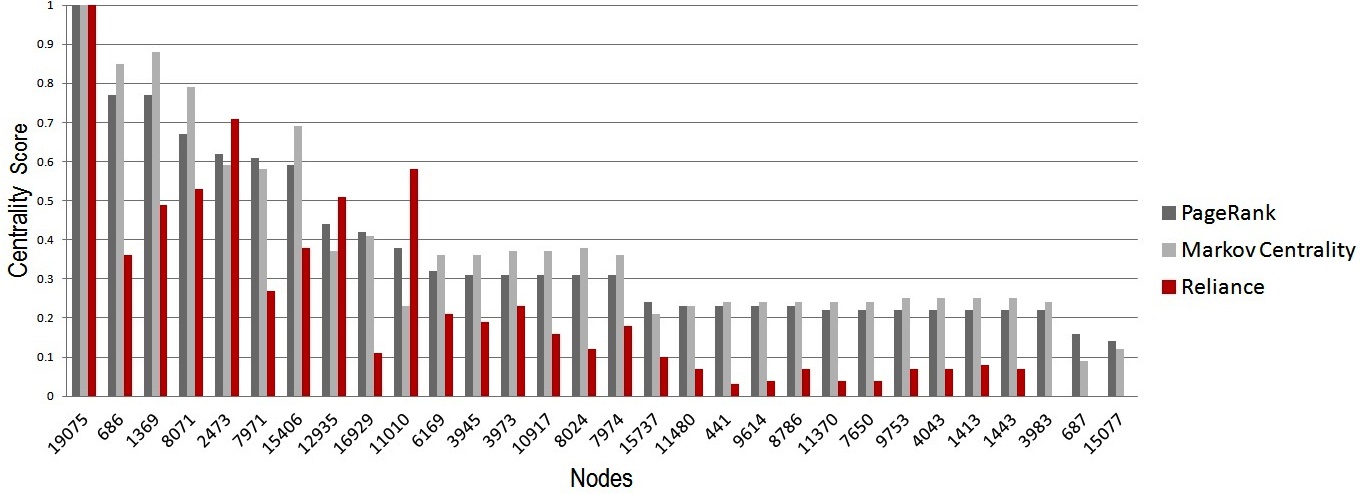}
\captionsetup{singlelinecheck=off,justification=justified}
\caption{\small \textbf{The Enron finance manager BCC sub-network node ranking.} The total reliance of all finance managers on each $v$ in the Enron Finance Manager BCC sub-network is compared with the PageRank and Markov centrality value of $v$. The nodes are ranked based on the descending order of PageRank values. There is a clear difference in node ranking, with for example, nodes 3945, 3973, 10917 and 7974 being ranked similarly by Markov centrality and Pagerank method but very differently by the reliance measure.}
\label{fig:FMComMarkovandPageRank}
\end{figure}

The total reliance ranking of the intermediate nodes  $v$ is next compared with the other two ranking methods,  PageRank and Markov centrality, in the Enron Finance Manager BCC sub-network. Again the values are normalised by dividing each value by the largest value in the respective set. The results are shown in Figure \ref{fig:FMComMarkovandPageRank}. The nodes are ranked based on the descending order of PageRank scores.

The bar chart in Figure \ref{fig:FMComMarkovandPageRank} shows the differences in node ranking between PageRank,  Markov centrality and the reliance measure. A few interesting differences are visible in this figure. Nodes 3945, 3973, 10917 and 7974 are ranked very differently by reliance but similarly by PageRank and Markov centrality. In contrast, nodes 3983, 687 and 15077 are valued by Markov centrality and PageRank but not relied on at all by the finance managers (see figure \ref{fig:FMComMarkovandPageRank} towards the right of the bar chart). Finally, 11010 (Lea Fastow) was also not picked out by either PageRank or Markov centrality. 

The comparisons are repeated for all the Enron managers, a larger algorithm feed, to see again if there is a difference in the ranking. A shortest paths sub-network is formed using all the managers and named the Enron Manager BCC sub-network. This sub-network is an bigger undirected graph with 121 nodes and 314 edges. First the comparison is carried out between the reliance measure and the betweenness measures, as before, and the intermediate nodes are ranked based on descending order of betweenness centrality (Brandes) values. 

\begin{figure}[ht]
\DeclareGraphicsExtensions{.pdf,.png,.jpg,.tiff}
\centering
\includegraphics[height=1.8 in,width=3.5 in]{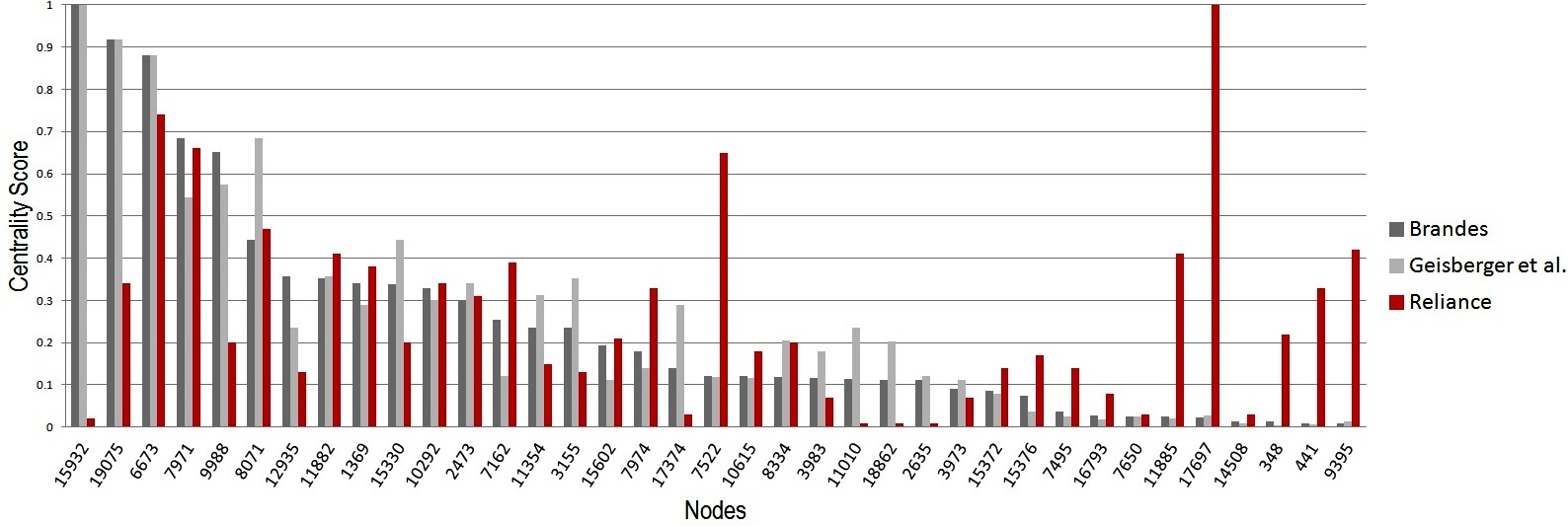}
\captionsetup{singlelinecheck=off,justification=justified}
\caption{\small \textbf{The Enron manager BCC sub-network node ranking.} 
 The total reliance of all managers on each $v$ in the Enron Manager BCC sub-network is compared with the betweenness centrality measures of Brandes, and Geisberger et al., with rankings ordered in descending order of Brandes' betweenness. Note that only the nodes with positive reliance value are displayed along the $X$-axis.}
\label{fig:BCValueM}
\end{figure}

Even more than in Figure \ref{fig:BCValueFM}, Figure \ref{fig:BCValueM} shows the difference in the ranking between the betweenness centrality measures and the reliance measure; the node (17697) picked as the most important one by the reliance measure is not picked by either Brandes or Geisberger et. al, both of which pick 15932, a node regarded as irrelevant by the reliance measure. Moreover, the betweenness centrality values for nodes 348, 441 and 9395 are almost the same (and negligible) whereas the reliance measure shows a different ranking with much higher values. 

\begin{figure}[ht]
\DeclareGraphicsExtensions{.pdf,.png,.jpg,.tiff}
\centering
\includegraphics[height=1.8 in,width =3.5 in]{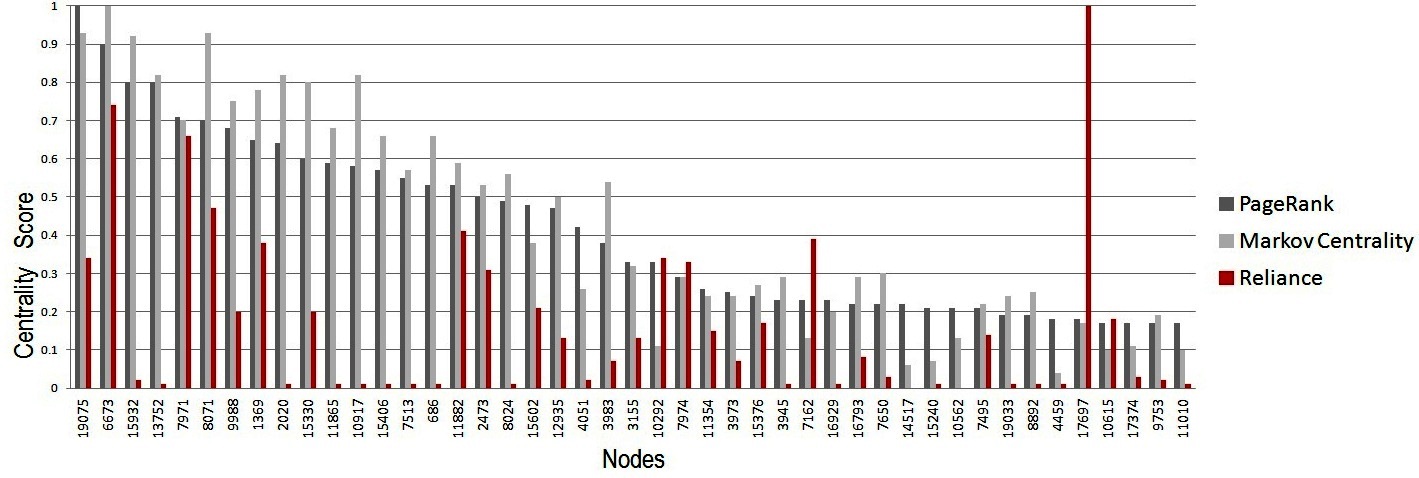}
\captionsetup{singlelinecheck=off,justification=justified}
\caption{\small \textbf{The Enron manager BCC sub-network node ranking.} 
The total reliance of all managers on each intermediate $v$ in the Enron Manager BCC sub-network is used for this comparison. The intermediate nodes (on the $X$-axis) are ranked based on the descending order of PageRank values. There is a clear difference in the ranking of the nodes using the three different methods. For example PageRank, Markov centrality and the reliance measure rank nodes 19075, 6673 and 17697 as having the highest value respectively.}
\label{fig:MComMarkovandPageRank}
\end{figure}

Finally we compare intermediate node reliance with PageRank and Markov centrality in the Enron Manager BCC sub-network. The result is shown in Figure \ref{fig:MComMarkovandPageRank}. The nodes are ranked based on the descending order of PageRank values. Again there is a clear difference between the three rankings.

From the comparisons done above, we can see that an investigation using the reliance measure rather than any of betweenness centrality, PageRank or Markov centrality rankings, will lead to  different, possibly more relevant, people to investigate. For example,  Lea Fastow was not in the algorithm feed for the experiment \cite{MagalingamP2014} using either all managers or finance managers, however, as illustrated in  Figure \ref{fig:BCValueFM},  Lea Fastow, the wife of Andrew Fastow \cite{Kathleen2003} and a convicted money laundering criminal \cite{Kathleen2003}, was heavily relied on by the finance managers. 

\subsection{Comparing reliance node ranking with betweenness centrality, Markov centrality and PageRank in the Noordin Top terrorist network}
\label{CompareBCandRelianceonNoordinTopTerroristNetwork}

\begin{figure}[ht]
	\centering
	\subfigure[\small The terrorist-friendship sub-network nodes ranked using betweenness centrality by Brandes, Geisberger et al. and reliance.]
	{
	\includegraphics[height=2.0in, width=3.4in]{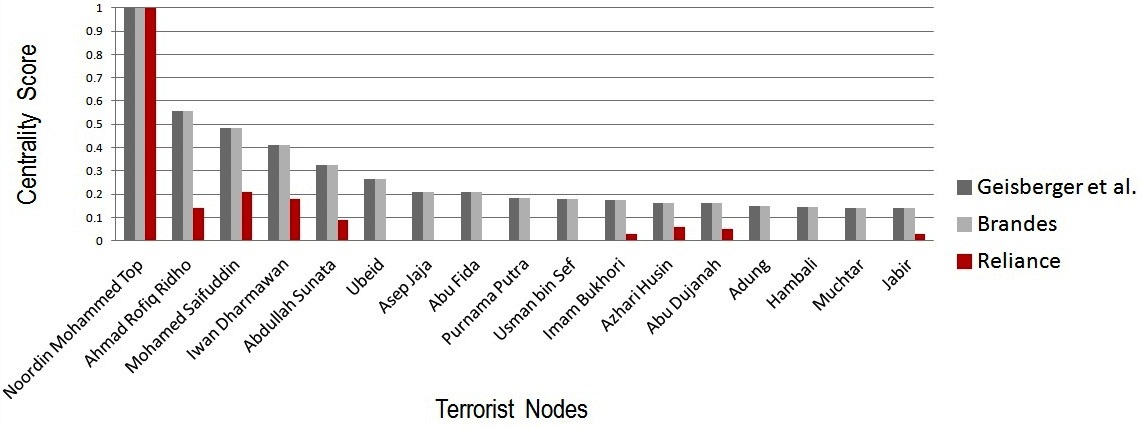}
	\label{fig:first_sub}
	}
	\\
	\subfigure[\small The terrorist-classmate sub-network nodes ranked using reliance, betweenness centrality by Brandes and Geisberger et al. ]
	{
		\includegraphics[height=2.0in, width=3.4in]{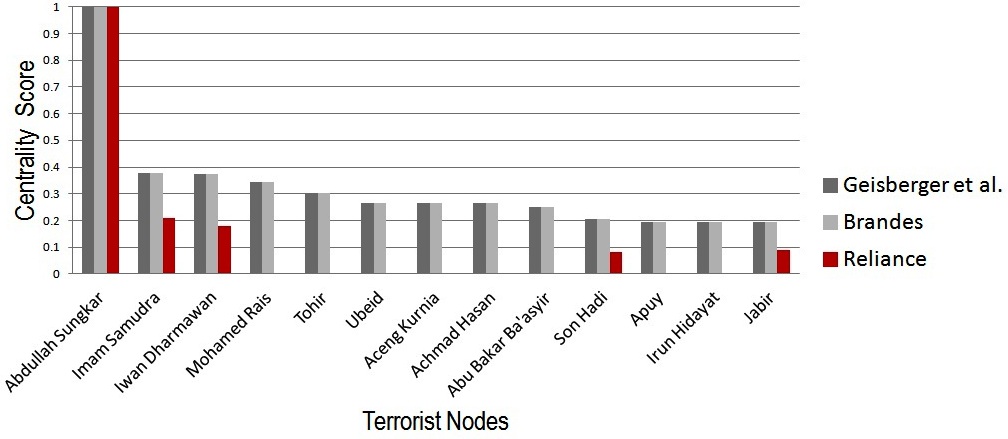}
		\label{fig:second_sub}
	}
\captionsetup{singlelinecheck=off,justification=justified}
	\caption{\small The nodes $v$ are listed on the $X$-axis, with the Brandes and Geiseberger et. al and reliance values obtained by each node depicted as the height of the bars, with the nodes ranked in descending order of Brandes' betweenness centrality values.  For each graph, nodes with zero reliability and with a betweenness (Brandes) value below that of the node with the lowest reliability, are not included.}
\label{fig:TerroristBetweennessRelianceCompre}
\end{figure}

\begin{figure}[ht]
	\centering   
	\subfigure[\small The terrorist-friendship sub-network nodes ranked using reliance, PageRank and Markov centrality.]
	{
		\includegraphics[height=2.0in, width=3.4in]{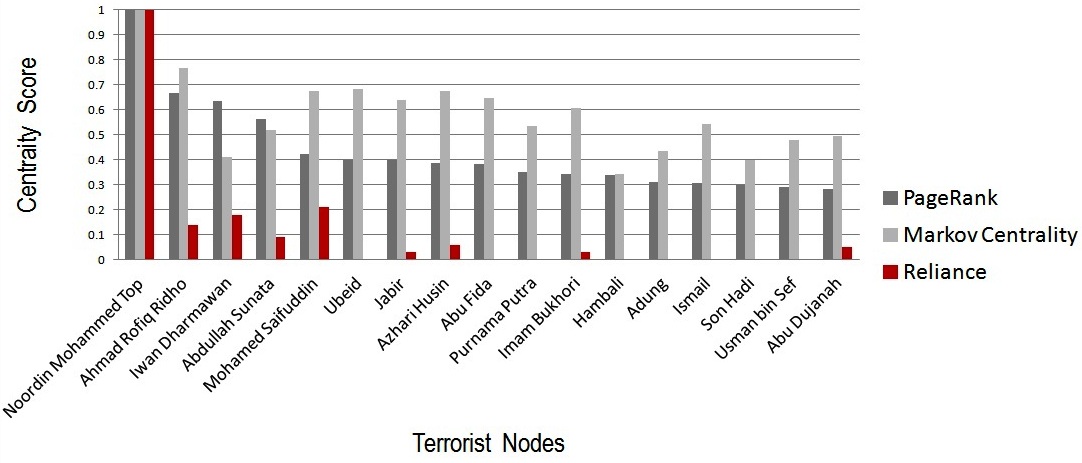}
		\label{fig:third_sub}
	}
    \\
	\subfigure[\small The terrorist-classmate sub-network nodes ranked using Markov centrality, PageRank and reliance.]
	{
		\includegraphics[height=2.1in, width=3.4in]{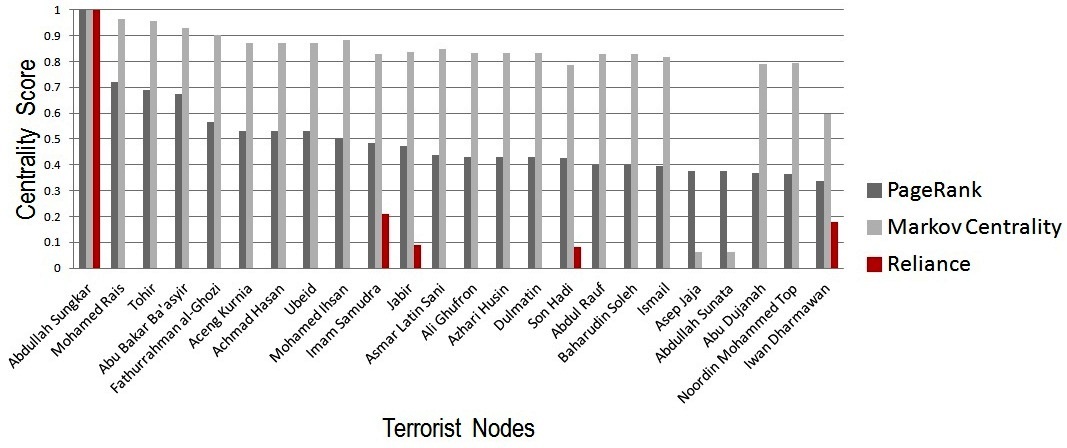}
		\label{fig:fourth_sub}
	}
     \captionsetup{singlelinecheck=off,justification=justified}
	\caption{\small The nodes $v$ are listed on the $X$-axis, with the PageRank, Markov and reliance values obtained by each node depicted as the height of the bars, with the nodes  ranked in descending order of PageRank scores.  For each graph, terrorists with zero reliance and ranked by PageRank scores as below the terrorist with the lowest reliance, are not included.}
\label{fig:TerroristPageRankMarkovRelianceCompre}
\end{figure}

Four different bar charts are presented here showing the comparison of intermediate node reliance with  betweenness, Markov centrality, and PageRank. The terrorist-friendship sub-network consists of 61 nodes and 91 edges and the terrorist-classmate sub-network has 39 nodes and 175 edges. We first isolate the highest reliance of each node on an intermediate node in the shortest paths between the former node and all other nodes in the sub-network. We then sum over all the reliance values for the same intermediate node and normalise it with the highest value in the list. We do a similar normalisation for the betweenness as well as PageRank and Markov centrality values.The results are shown in Figures \ref{fig:TerroristBetweennessRelianceCompre} and \ref{fig:TerroristPageRankMarkovRelianceCompre}.

\vspace{0mm}

Both Figures \ref{fig:TerroristBetweennessRelianceCompre} and \ref{fig:TerroristPageRankMarkovRelianceCompre}, illustrate the versatility of the reliance measure, with very few nodes showing up as being relied on by all the members of the group. Thus, for example, taking Figure \ref{fig:TerroristPageRankMarkovRelianceCompre}(b), an investigator could narrow their first detailed investigations to just 5 members of the group.

\section{Identifying crime priority nodes using the reliance measure}
\label{IdentifyingCrimePriorityNodesMain}

The experiment hereafter uses only the Enron dataset and the reliance measure to get the persons of interest (a.k.a. ``crime priority nodes") of each finance manager and manager in both the `BCC' and `TO/CC' network respectively.  The main purpose of these experiments is to identify the intermediate nodes important to the suspects, as ranked by the reliance measure. At this point, we do not corroborate the identified persons of interest with any published articles or past research.

\subsection{Extracting the suspects' crime priority nodes from the Enron `BCC' email network}
\label{suspectssub-networkBCC}

We first show the results of identifying the Enron money laundering suspects' important nodes in the `BCC' email network based on suspect-intermediate reliance value. We start by choosing the Enron finance managers as our source nodes, and extract the sub-snetwork using SPNSA. The  reliance of each finance manager on every intermediate node ($v$) in the path between that finance manager and all other nodes within the  Enron Finance Manager BCC shortest paths sub-network, is calculated. The intermediate node that each finance manager relies on the most is identified. The list of these crime-priority nodes reveals 2 money laundering criminals (\cite{MagalingamP2014, Kathleen2003}) as important to each other; Ben Glisan (1369) and Andrew Fastow (686). 

Next, we use all managers (including the finance managers) as the suspects (SPNSA algorithm feed), to see if  new people of interest are obbtained. Unlike the Enron Finance Manager BCC sub-network,  there is no criminal to criminal reliance in this network. However, nodes that are the persons of interest to the managers include managers Greg Whalley (6673), President and Chief Operation Officer, and Lou Pai (11357), CEO of Enron Energy Services.

\subsection{Extracting suspect's crime priority nodes from the Enron `TO/CC' email network}
\label{suspectssub-networkTOCC}

Using the same method as shown above, we extract the finance managers' crime priority nodes from the `TO/CC' shortest paths sub-network. Note that some employees' email addresses that exist in the `BCC' network do not occur in the `TO/CC' network, for example email address (andrew.fastow@ljminvestments.com) belonging to Andrew Fastow occurs only in the `BCC' network. Thus, different IDs were used in the `TO/CC' network. 

The finance managers \cite{anderson2003}  used as the SPNSA algorithm feed are Andrew Fastow (1472), Sherron Watkins (26577), Ben Glisan (2521), Rick Causey (24277) and Jeff McMahon (12919). George Wasaff (george.wasaff@enron.com (10351)) is the intermediate node that all finance managers rely on the most in this network, which is an interesting fact and could be the starting point for further investigation. 



It is possible that more criminal to criminal connections would be identified if an investigator picked the best possible suspects to inspect (based on corroborative information or interviews). The experiments in this section show that the reliance measure allows an investigator to find people who are close to and heavily relied on by the suspects for further investigation.

\section{Conclusion}
\label{DiscussionforCaseStudy1}
This paper introduced a new reliance measure to rank nodes in a network and therefore identify nodes of interest. This reliance measure is different from betweenness centrality because the latter calculates the centrality value of a node for all sources whereas the former calculates the reliance of a given list of specific sources on a node. We compared the reliance ranking with other centrality measures, Google PageRank and Markov centrality, as well as betweenness centrality. Reliance identified a very different subset of nodes from those identified by the other measures. We showed that the ranking based on the reliance measure could also be used to identify the nodes relied on the most by particular persons of interest to an investigation.

Our SPNSA as described in \cite{MagalingamP2014} is able to produce a small and manageable network. In this paper we furthered our research and analysed the connections between nodes in the network; reliance of one node on another leading possibly to the identification of important nodes. This reliance method could reduce the time needed to explore a large network and hence may speed up an investigation process. It is important to note that, prior to the application of the shortest paths network search algorithm \cite{MagalingamP2014} and the reliance formula, it is essential to choose the most relevant suspects to a crime. The analysis proposed in this paper could also yield more criminal to criminal connections by choosing the most applicable or appropriate algorithm feed for a crime incident. 



\bibliographystyle{IEEEtranS}

\end{document}